# Vacuum-Gap Capacitors for Low-Loss Superconducting Resonant Circuits

Katarina Cicak, Michael S. Allman, Joshua A. Strong, Kevin D. Osborn, and Raymond W. Simmonds

***Abstract*—Low-loss microwave components are used in many superconducting resonant circuits from multiplexed readouts of low-temperature detector arrays to quantum bits. Two-level system defects in amorphous dielectric materials cause excess energy loss. In an effort to improve capacitor components, we have used optical lithography and micromachining techniques to develop superconducting parallel-plate capacitors in which lossy dielectrics are replaced by vacuum gaps. Resonance measurements at 50 mK on lumped LC circuits that incorporate these vacuum-gap capacitors (VGCs) reveal loss tangents at low powers as low as $4\times10^{-5}$, significantly lower than with capacitors using amorphous dielectrics. VGCs are structurally robust, small, and easily scaled to capacitance values above 100 pF.**

## I. INTRODUCTION

HIGH-quality-factor superconducting resonators with low-loss microwave circuit components are used to build multiplexed arrays of many types of low-temperature detectors, such as superconducting transition-edge sensors (TES) [1], and microwave kinetic inductance detectors (MKID) [2]. In recent years there has been a push to engineer lossless components for superconducting quantum bits (qubits) as well. Phase qubits [3] incorporate a small Josephson junction that is often shunted by a low-loss capacitor to lower the qubit frequency into a manageable range of several gigahertz [4]. Undesirable coupling between superconducting qubits and spurious two-level system (TLS) defects found in amorphous dielectric materials on chip (such as $SiO_2$ or $SiN_x$ insulating spacers, or $AlO_x$ in the tunnel junction) disrupt the coherent properties of qubits [5]-[7]. Utilizing vacuum instead of dielectrics in circuit components can provide enormous benefits by mitigating effects of TLS.

Microfabricated planar capacitors with interdigitated fingers often used in superconducting qubits and other on-chip resonant circuits exhibit low microwave loss characteristics when fabricated on low-loss substrates such as crystalline sapphire (see for example Fig. 6 and [8]). However, these interdigitated capacitors (IDCs) have large footprints and take up valuable chip space. In addition, the long distributed fingers of IDCs contribute a significant amount of stray inductance, producing non-ideal frequency-dependent capacitors. Motivated by the need for lossless cryogenic microwave components and by the previous work on bulk low-loss capacitors [9], and considering the shortcomings of IDCs and dissipation in dielectric-filled capacitors, we set out to microfabricate a superconducting parallel-plate capacitor in which the dielectric material is replaced by vacuum. In such a vacuum-gap capacitor (VGC), the parallel-plate configuration with a sufficiently small gap can provide a more ideal capacitor behavior with a relatively small on-chip footprint, while completely confining electric fields to lossless media: superconducting electrodes and vacuum. Here we present the fabrication method and the loss characterization of VGCs.

## II. FABRICATION PROCESS

### A. VGC with Si-posts

By using standard optical lithography and bulk micromachining techniques [10], [11], we have developed a process to fabricate parallel-plate capacitors with a vacuum gap between the plates. The complete process is summarized in Fig. 1. The final step, shown in Fig. 1(e), is the most delicate part of processing. Here we remove a large fraction of the Si film, creating a parallel-plate capacitor with mostly empty space between the plates. A small amount of Si is left within the capacitor to form posts that support the suspended Al top plate electrode. To create this structure, a high-pressure $SF_6$ plasma etch is used to remove most of the sacrificial Si layer through an array of holes patterned in the top plate. The high pressure etch recipe ($SF_6$ gas flow: 100 sccm, pressure: 26.7 Pa (200 mTorr), power: 100 W, bias voltage: -20 V) provides a more isotropic removal of Si than typical lower-pressure recipes designed to produce a more directional vertical etch. Here the etching of Si extends laterally under the top Al plate, creating a circular undercut centered under each top-plate hole. As the etching progresses, the undercut circles grow until eventually they begin to intersect. The etching process is terminated when the remaining Si forms small post structures. Each post has a star-like cross-section (shown in Fig. 2) and is located under the center of each honeycomb formed by the holes in the top plate. Posts are thus arranged in a hexagonal lattice between the two plates of the capacitor.

Voltage applied across the VGC produces an electric field

Manuscript received 26 August 2008. This work was supported in part by IARPA under Grant No. DNI-713268.

Katarina Cicak, Michael S. Allman, Joshua A. Strong, and Raymond W. Simmonds are with the National Institute of Standards and Technology, Boulder, CO 80305 USA (phone: 303-497-4606; fax: 303-497-3042; e-mail: cicak@boulder.nist.gov).

Kevin D. Osborn was with the National Institute of Standards and Technology, Boulder, CO 80305 USA. He is currently with the Laboratory for Physical Sciences, College Park, MD 20740, USA.

between the plates that is confined mostly within the vacuum gap with some residual field left within the Si posts. To ensure low loss, it is desirable to maximize the vacuum volume while minimizing the volume occupied by posts. However, the posts must be robust enough to provide sufficient structural support. Achieving the proper etch time is crucial but complicated in practice, due to undercut etch rates that: (1) are not constant (i.e., the etching slows with time as the Si being etched is farther and farther away from the holes in the top plate), (2) are not uniform across the wafer, due to size and shape variation of the holes in the top plate, and (3) change from wafer to wafer depending on the precise Si-film thickness and the condition of the etching chamber at the time of processing.

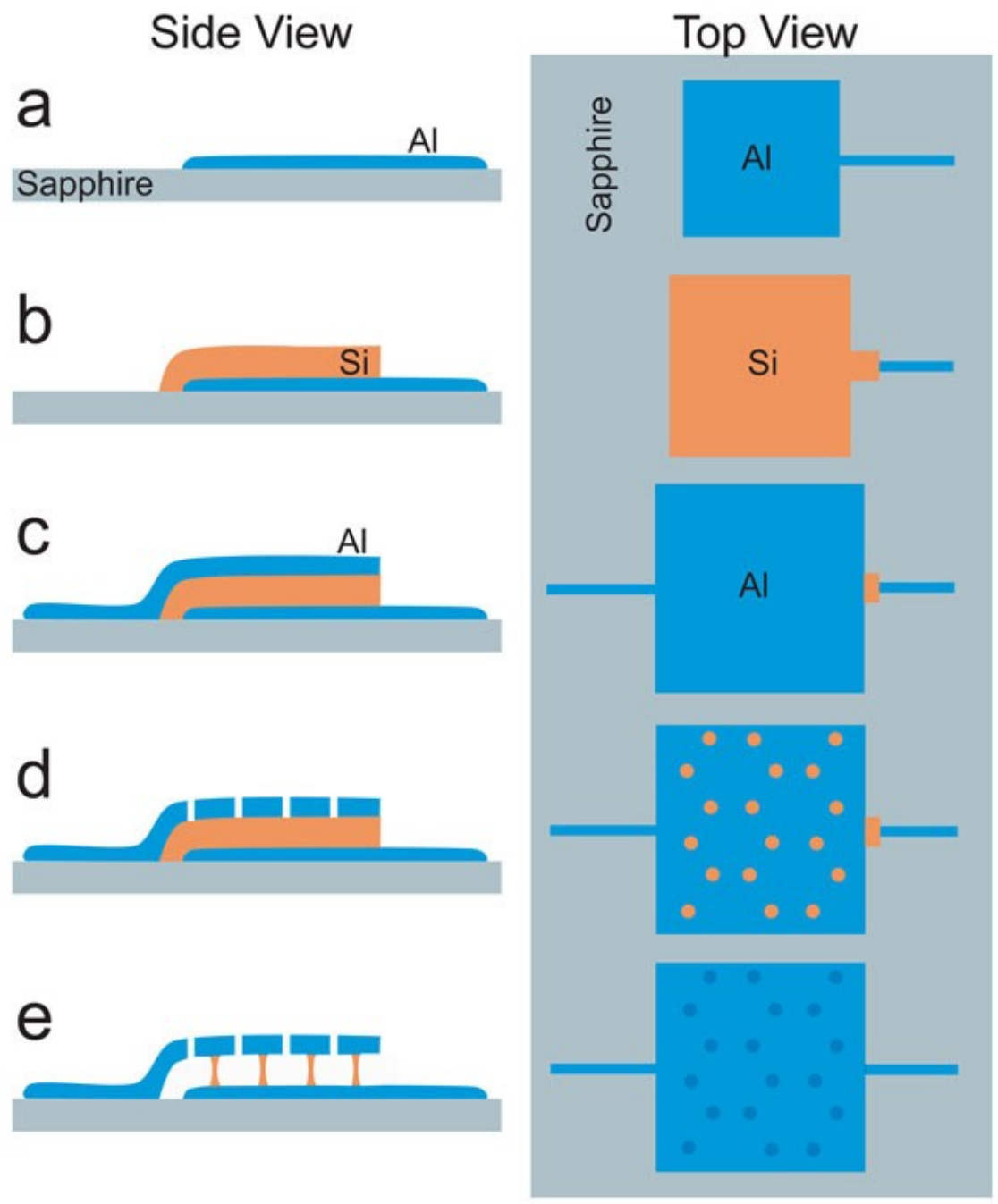

Fig. 1. Fabrication procedure for VGCs with Si-posts. (a) A 100 nm layer of amorphous Al is sputtered on a 3 inch sapphire wafer and patterned by wet-etching to form a base plate of a capacitor. (b) 200 nm of amorphous Si is deposited by rf-sputtering and patterned by $SF_6$ plasma etch to cover the base capacitor plate. (c) A second Al layer is deposited and patterned to form a top capacitor plate. (d) A pattern of small holes 1 μm in diameter and arranged in a honeycomb lattice, is patterned in the top plate by wet-etching Al, exposing the Si-layer below. (e) The wafer is diced into individual device chips, and then all the chips are etched at the same time in a high-pressure $SF_6$ plasma that removes most of the Si through the top-plate holes. The etch process is terminated before all Si is removed. The remaining Si forms an array of small support posts that prevent top electrode from collapsing. It is important that the wafer be diced *before* this last etch step. The cooling water sprayed over the wafer during dicing, or a wet process to remove a protective resist layer after dicing, would collapse the capacitor if the sacrificial layer had already been removed. If VGCs are to be integrated with other components on the chip, the fabrication of these components should be completed before the wafer is diced. The components then must withstand the sacrificial $SF_6$ etch. Usually, this is not an overly restrictive fabrication requirement.

The total etch time to produce lateral undercut of ~8 μm in a Si layer 200 nm thick can vary from 15 to 25 minutes. To avoid over-etching the posts we remove Si in *several* short etch steps, and rely on inspecting chips under an optical microscope after each step. Since the posts are hidden under the Al top plate, we must resort to destroying several device chips by removing the top plate to inspect the posts below. The top plate can be easily removed by gently pressing a piece of adhesive polyimide tape over the chip surface and then removing it. The tape peels off the suspended Al top plate, leaving the Si posts intact for inspection. Alternatively, we can wet-etch all the Al away leaving the Si posts, as seen in Fig. 2(b)-(d). Another (destructive) way to reveal the posts is to force the soft Al film of the top plate to drape over the posts like a blanket over an array of rocks. This is accomplished by applying a drop of isopropyl alcohol over the chip. The surface tension of the drying alcohol pulls the two plates together, collapsing the top plate so that the posts appear as bumps and are readily visible under a microscope, as shown in Fig. 2(a). By careful engineering we have achieved vacuum-volume to post-volume ratios as high as 0.99 with post widths $w < 1$ μm, and with typical widths in the 1-3 μm range.

We have explored geometries with different distances between posts and found that a 10 μm nearest-neighbor distance between holes in the top plate, which produces a post-to-post distance of 17.3 μm, yields stable, robust VGCs. We have also produced VGCs with plate separations, *d*, of 200 nm and 500 nm. In order to maintain a large capacitance with a small footprint, most of the measurements reported in this paper were performed on capacitors with $d = 200$ nm.

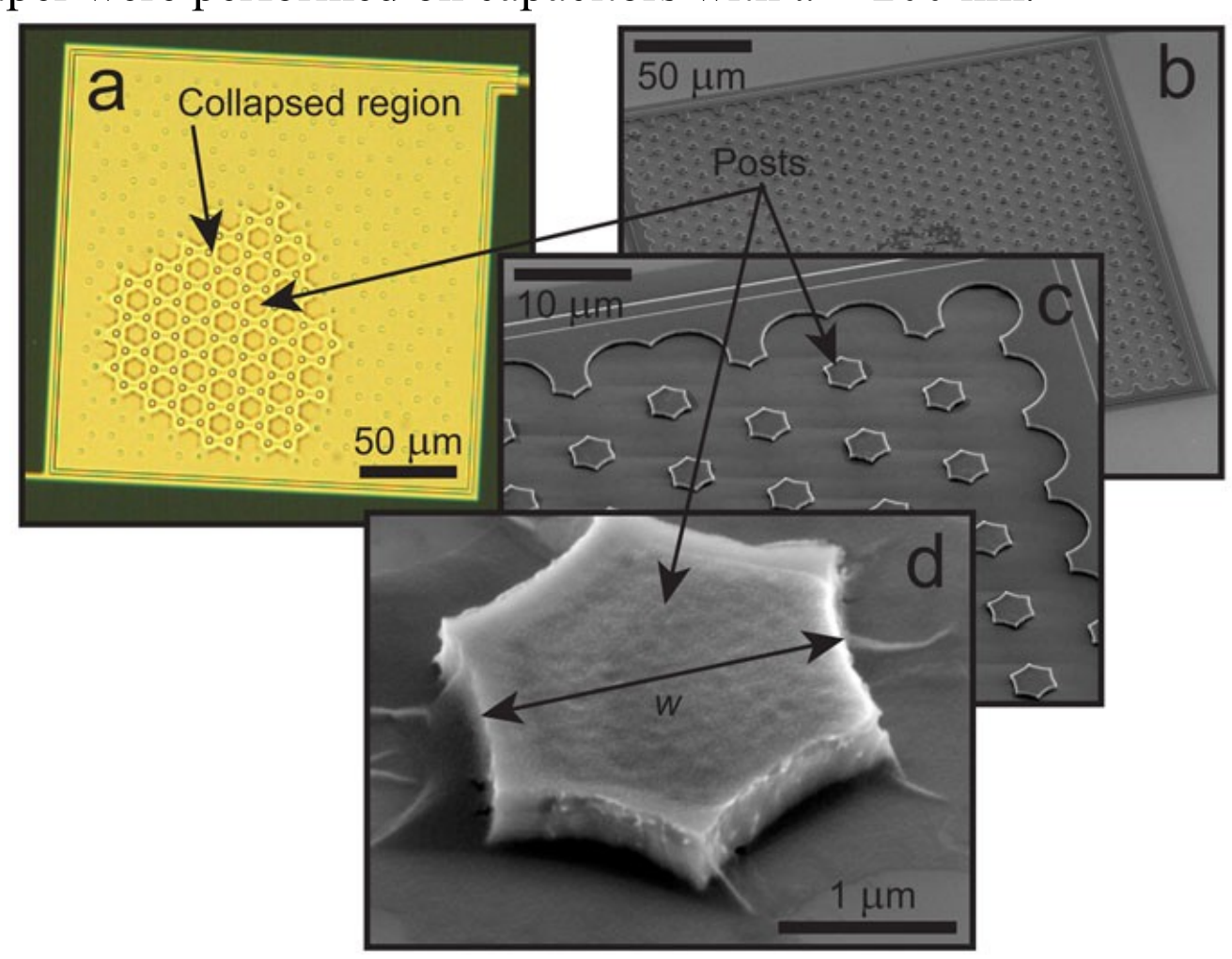

Fig. 2. (a) Micrograph of a VGC with Si posts partially collapsed by isopropyl alcohol. Large posts under top plate are visible in the collapsed region. (b)-(d) show SEM images of remnants of another capacitor in progressive magnifications. In these images all aluminum has been removed by wet-etching, leaving behind a hexagonal array of Si posts.

### B. *VGC with Si-posts in base holes*

In an attempt to further remove the residual electric fields from the Si posts, we modified our original VGC design by eliminating the Al base plate under and in the vicinity of each post. This is accomplished by patterning holes in the base plate after the first step in the fabrication (Fig. 1(a)), in places where the posts will form during the subsequent $SF_6$ plasma etch. Choosing a base-hole diameter, $2r$, larger than the final post width, $w$, assures that the posts rest directly on the sapphire substrate (Fig 3(b)). For a given applied voltage across the capacitor plates, the amplitude of the electric field within a small Si post ($w \ll 2r$) is minimized in the limit of $r/d \rightarrow \infty$. In our fabricated VGCs with base holes, $r/d \sim 10$–30, so that the residual electric fields within the posts are expected to be comparatively small. A fabricated VGC with base holes is shown in Fig. 4.

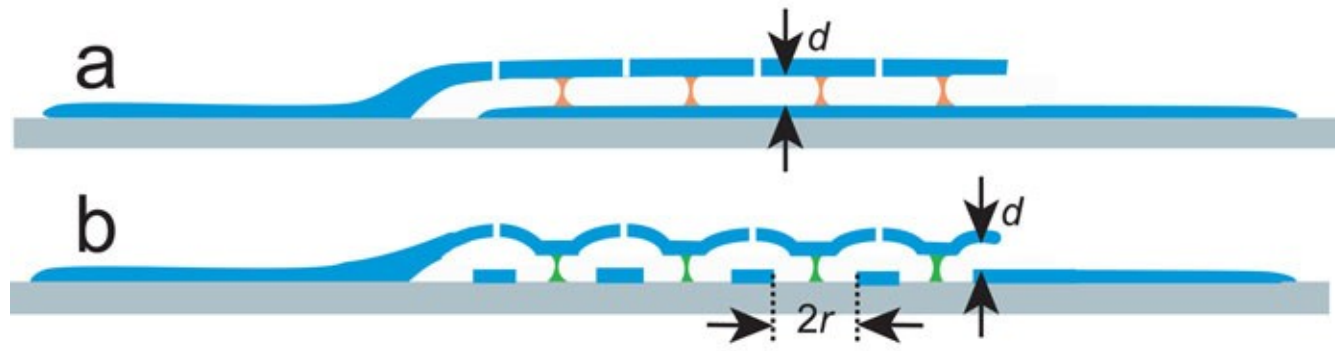

Fig. 3. (a) Original VGC design. Electrodes *above* and *below* posts: electric fields confined within posts and vacuum. (b) VGC with base holes. No electrode below posts: electric fields predominantly in vacuum. Si posts can be replaced by superconducting Nb posts.

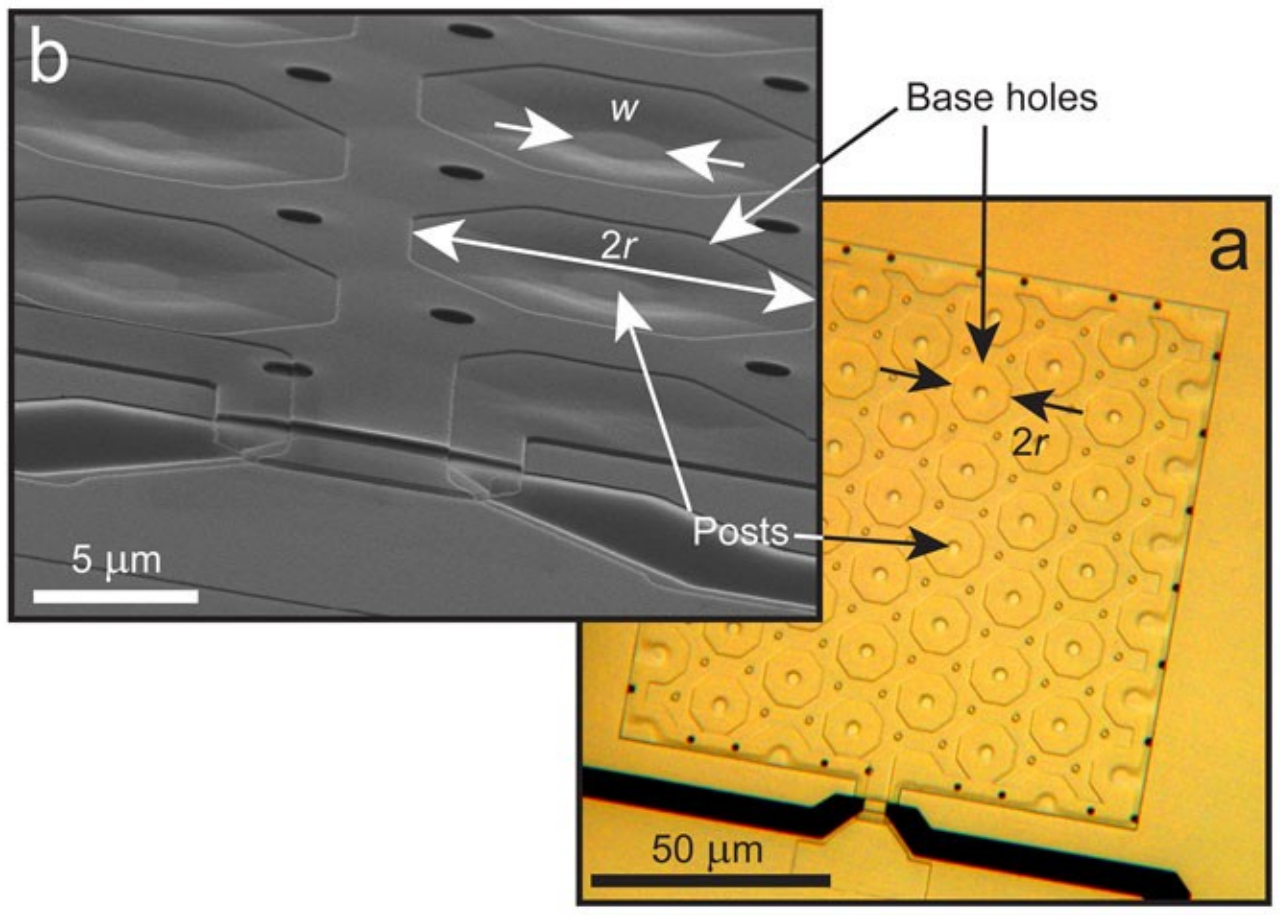

Fig. 4. (a) Micrograph of a VGC with base holes collapsed by isopropyl alcohol. (b) SEM image of a VGC with base holes. Electron beam from the SEM charges both plates negatively, so they repel each other resulting in the top plate slightly bulging away from the bottom one.

### C. VGC with Nb-posts in base holes

In the new VGC design with base holes, the posts provide only structural support, but do not electrically separate the two plates. To further eliminate lossy dielectrics, we replaced the Si posts with superconducting Nb posts. The fabrication requires only the substitution of Nb instead of Si in the step shown in Fig. 1(b). The posts can still be created with a high-pressure $SF_6$ plasma etch (Fig. 1(e)), although the undercut etch rates for Nb are much lower than those for Si.

## III. Loss Measurement Configuration

To evaluate the microwave loss characteristic of our VGCs, we incorporate them in simple lumped-element LC circuits and measure the quality factors of these resonators. As shown in Fig. 5, a capacitance $C$ (provided by two VGCs each with $C/2$ in a range 0.6-1.6 pF) is placed in parallel with a loop of Al wire of inductance $L \approx 360$ pH. Two small planar IDCs, each with a capacitance $C_C$ ($\approx$ 1.5, 3.8, or 7.6 fF $\ll C$), provide coupling to the resonator via two transmission line ports with characteristic impedance $Z_0 \approx 50\ \Omega$. The chip is wire-bonded to a shielded circuit board that is cooled in vacuum to 50 mK using a dilution refrigerator. The resonator is driven by a voltage signal from a synthesized signal generator through the input port, and the transmitted power is measured at the output port with a spectrum analyzer after a circulator and a cryogenic (4 K) amplifier.

The circuit schematic of the resonator is shown in the inset of Fig. 5. The loss in the resonator can be modeled by an effective resistive component $R$ in parallel with the LC components. On resonance the impedance of the shunting LC circuit approaches ∞, and the frequency spectrum exhibits a peak in transmission with a Lorentzian line shape at

$$f = f_R \approx (2\pi)^{-1}(LC)^{-1/2} \tag{1}$$

for $C_C \ll C$. $R$ is related to the characteristic loss tangent $\tan\delta$ through

$$\tan\delta = 1/Q_I = (2\pi f_R RC)^{-1} \approx R^{-1}(L/C)^{1/2}, \tag{2}$$

where $Q_I$ is the internal quality factor of the resonator. $Q_I$ can be extracted from the relation

$$Q_M^{-1} = Q_I^{-1} + Q_C^{-1}, \tag{3}$$

where $Q_C \approx (C/C_C)/(4\pi f_R Z_0 C_C)$ for $C_C \ll C$ represents the loss from coupling to the external feed lines, and $Q_M$ is the measured quality factor obtained from $Q_M = f_R/f_{FWHM}$ where $f_{FWHM}$ is the full width at half maximum of the resonant peak.

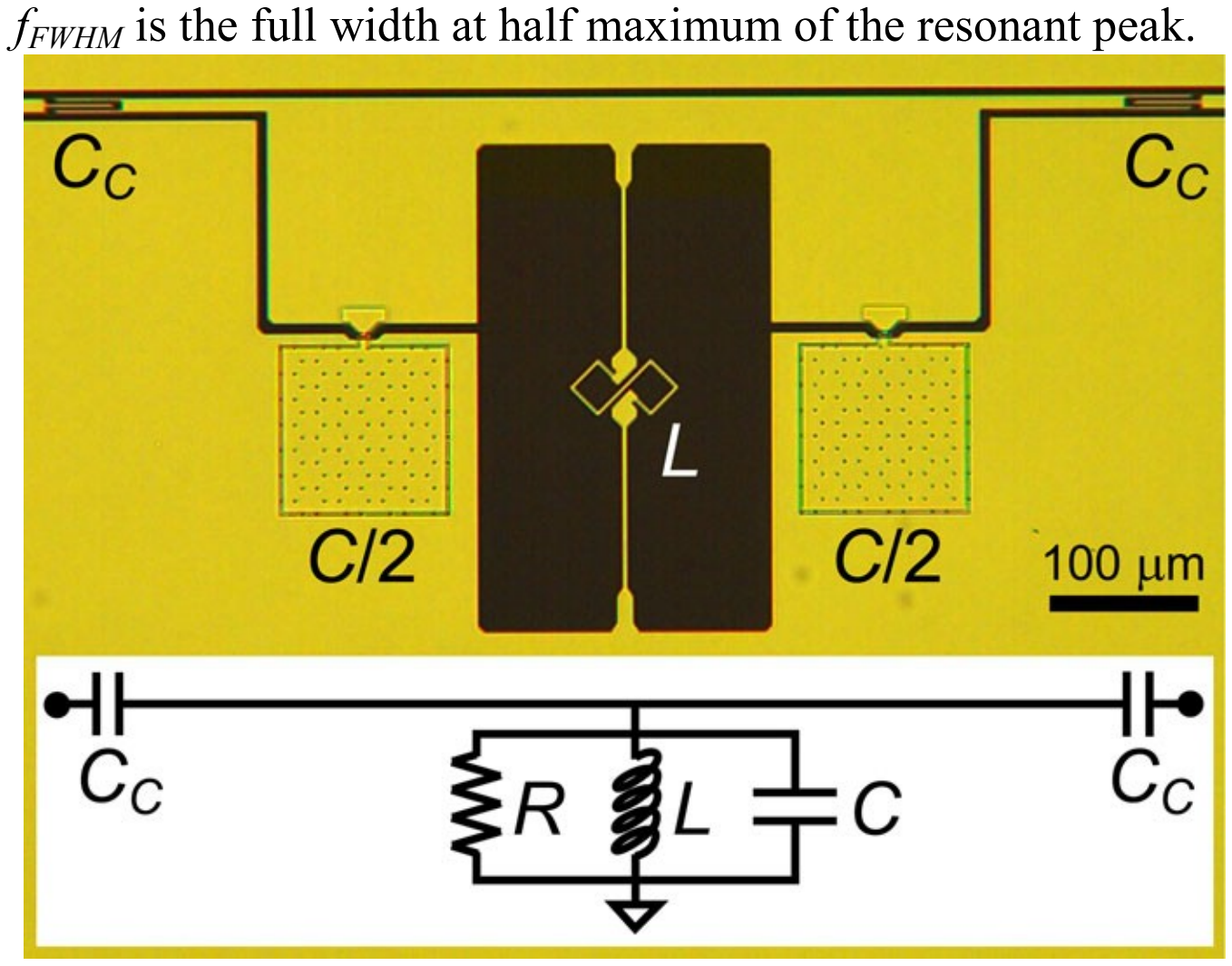

Fig. 5. Fabricated LC resonator with the inductance $L \approx 360$ pH. The capacitance is split into two identical VGCs for symmetry, with measured $C/2 \approx 1.2$ pF. $C_C \approx 3.8$ fF. The inset shows the equivalent circuit schematic.

## IV. Results and Analysis

Transmitted power spectra are measured in the 4-8 GHz range on resonators with various capacitors, and $\tan\delta$ is extracted as discussed in the previous section. Fig. 6 shows typical loss-tangent results as a function of the RMS voltage across the lumped LC-circuit obtained from resonators using VGCs with Si posts, with and without base holes, and a VGC with Nb-posts in base holes. For comparison, we include our loss data from three amorphous dielectric materials commonly used in parallel-plate capacitors. We also include data we obtained from a planar IDC on a sapphire substrate with a footprint roughly ten times larger than that of an equivalent VGC. Due to the absorption of energy by parasitic TLSs in the dielectric materials, which become saturated at high powers [5], the observed loss tangents are large at low voltages and decrease with increasing voltage across the resonator. This effect is stronger for the capacitors filled with dielectric materials than for VGCs, confirming that the removal of dielectrics removes the dissipative TLS defects. At very low powers (in the single-photon limit, where phase qubits and other superconducting quantum devices operate), the VGC with Nb-posts achieves a loss tangent of ~4x10$^{-5}$, within a factor of two of the IDC value. Given the many shortcomings of IDCs discussed earlier, VGCs should be the capacitors of choice for many low-loss, compact superconducting circuits.

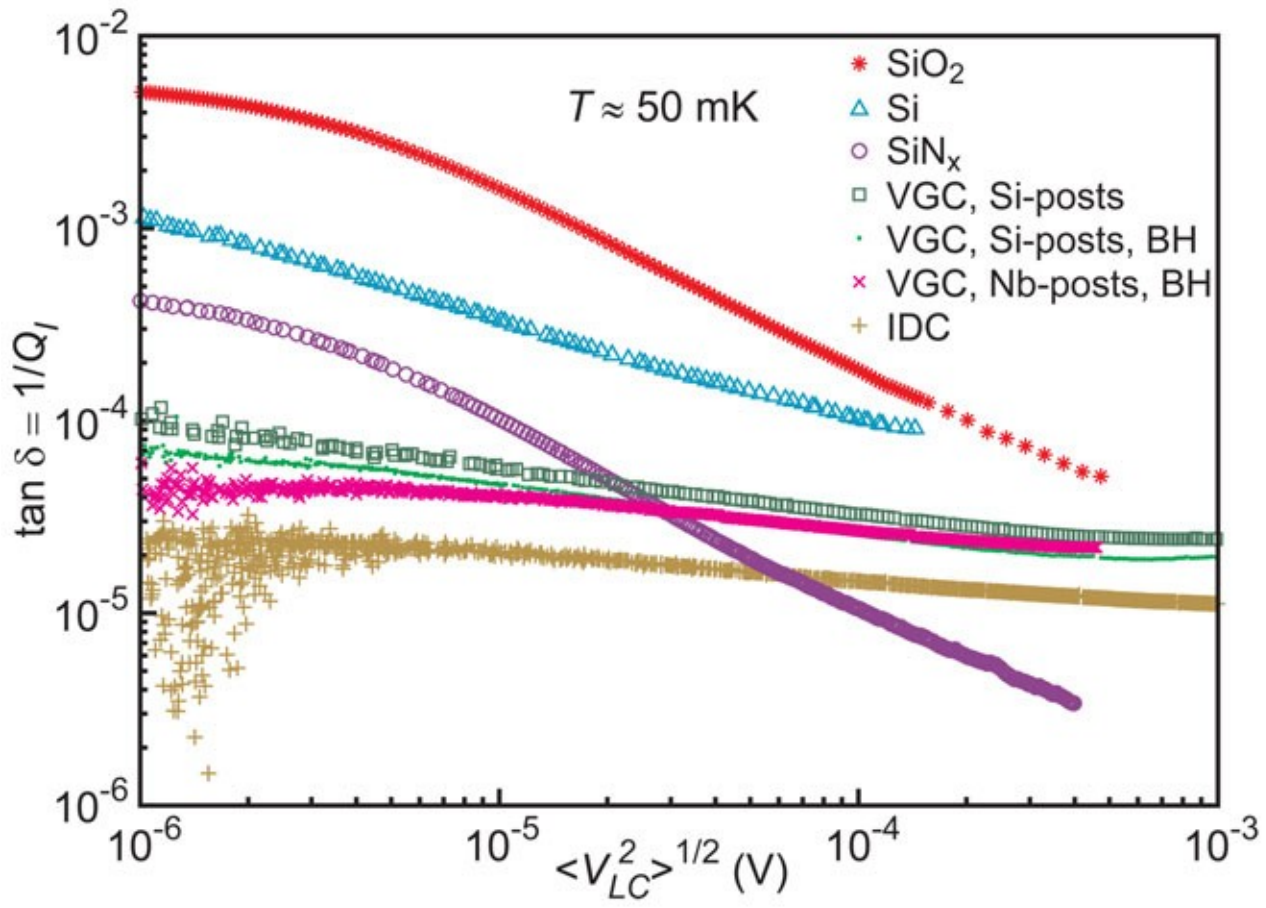

Fig. 6. Measured loss tangent vs. RMS voltage across the resonator, $<V_{LC}^2>^{1/2}$, for various capacitor configurations. BH stands for base holes. The $C$, $L$, and $C_C$ component values for each data set are as follows: $C \approx 1.69$, 2.67, 2.95, 2.48, 3.03, 1.39, 1.61 pF; $L \approx 360$, 360, 723, 360, 360, 360, 320 pH; $C_C \approx 15$, 7.6, 4.0, 3.8, 7.6, 1.5, 2.0 fF in the order as listed in the legend of the figure. The chosen values for $C_C$ produce large $Q_C$, so that $Q_I$ is close to $Q_M$ at low $<V_{LC}^2>^{1/2}$ values.

The residual loss in the best VGC may be associated with amorphous inner-surface native oxides on the superconducting Al plates [12]. We intend to test this in the future by probing how loss in VGC scales with $d$. For a larger $d$, a given voltage across the capacitor will produce a smaller electric field in the VGC. TLS defects in surface oxides are then expected to couple more weakly to the electric fields. Although the previous work clearly points to the TLS material defects as a source of loss [4]-[7] the details of the microscopic mechanisms are not yet fully understood.

We find that the capacitance $C$ obtained through (1) from the measured $f_R$ scales linearly with the total plate overlap area $A$ (with the area of the base holes excluded) of the VGCs. However, we find a systematic multiplicative factor between measured values $C$ and the design values $C_g = \varepsilon_0 A/d$ (where $\varepsilon_0$ is the permittivity of free space and $d$ is the expected plate separation). The ratio $C/C_g$ is summarized in Table 1 for different VGCs. The systematic factor from the VGC without base holes ($C/C_g = 2.2$) is slightly larger than what we would expect when we account for the dielectric contribution from the Si posts. In the case of VGCs with base holes, this factor is much larger than anticipated. We believe that the source of this factor derives from sagging of the top plate with respect to the bottom plate, providing an effective separation $d_{eff}$ smaller than the thickness $d$ of the sacrificial layer (i.e., post height). We can rule out electrostatic forces from stray floating charge on the plates as a possible cause of sagging, because the capacitor plates are shorted through the parallel inductor. The RMS voltage across the plates, which produces an attractive force when driving on resonance, cannot explain the magnitude of $C/C_g$. We verified that we see no significant frequency shift when measuring $f_R$ for low and high powers.

TABLE 1 SUMMARY OF VGCS USED IN LC MEASUREMETS

| Post type | Base hole diameter, $2r$ (μm) | $C$ per 100 μm × 100 μm footprint (pF) | $C/C_g$ |
|---|---|---|---|
| Si | 0 | 0.98 | 2.2 |
| Si | 3.7 | 1.23 | 2.9 ± 0.1 (2)[a] |
| Si | 12 | 0.88 | 3.6 |
| Nb | 12 | 0.58 | 2.4 ± 0.3 (5)[a] |

[a]The number in parentheses indicates the number of LC circuits on which measurements were performed. The uncertainty reported is the standard deviation from all measurements. Where parentheses are absent, only a single LC circuit was measured.

Our devices have survived multiple thermal cycling from room temperature to below 100 mK, demonstrating that VGCs are structurally robust and do not collapse under usual operating conditions. However, we observed that thermal cycling can increase $C$ by up to 10%. Sagging is most likely an artifact of the stresses in films produced during fabrication in conjunction with mismatches in thermal contractions during cooldown. If we make a reasonable assumption that sagging is more pronounced midway between posts than near each post, then most of the capacitance is associated with the regions that are far from the posts, accentuating the difference between VGCs with and without base holes, as well as those with larger base holes. Our data are consistent with this idea, since we observe that $C/C_g$ increases as we go from VGCs without base holes ($2r = 0$) to those with base holes with $2r = 3.7$ μm, and further to base holes with $2r = 12$ μm. We can partially attribute the difference in $C/C_g$ between VGCs with superconducting Nb posts and those with Si posts ($2r = 12$ μm) to the residual dielectric constant of the Si posts. We note, however, that stress in the suspended Al plate may depend on the sacrificial material used. This makes it difficult to provide a direct comparison of $C/C_g$ for devices using different post materials. Although the sagging effects produce different $C/C_g$ values in different VGC types, the small spread in $C/C_g$ from several identical devices (see Table 1) indicates that it is possible to design a capacitor with a predictable capacitance value if $C/C_g$ is known.

Preliminary measurements show that VGCs can be easily scaled up to produce capacitance as large as 100 pF. Obtaining such high values with large distributed structures laden with stray inductance such as IDCs is impractical on chip.

## V. CONCLUSION

We have successfully fabricated capacitors with vacuum between parallel plates. By removing dielectrics, we eliminate undesirable TLS defects, thus providing components with very low loss tangents. VGCs can find a wide range of use in any low-temperature circuit requiring a small on-chip footprint and low losses. They are ideal for applications in multiplexed detector readouts and superconducting quantum bits.